\documentclass{iau}

\usepackage{graphicx}

\title[Reionization Models Classifier] 
{Reionization Models Classifier using 21cm Map Deep Learning}

\author[Hassan et al.,]   
{Sultan Hassan$^{1,3}$, Adrian Liu$^2$, Saul Kohn$^3$, James E. Aguirre$^3$, Paul La Plante$^3$, Adam Lidz$^3$}

\affiliation{$^1$ The Department of Physics and Astronomy, \\ University of the Western Cape, \\ Bellville, Cape Town, 7535, South Africa \\ email: {\tt sultanier@gmail.com} \\[\affilskip]
$^2$ Department of Astronomy and Radio Astronomy Laboratory, \\ University of California Berkeley,
Berkeley, CA 94720, USA\\
$^3$ Department of Physics and Astronomy, University of Pennsylvania,\\
Philadelphia, PA 19104 }

\pubyear{2018}
\volume{333}  
\jname{Peering towards Cosmic Dawn}
\editors{Vibor Jeli\'c \& Thijs van der Hulst, eds.}
\begin{document}

\maketitle

\begin{abstract}
Next-generation 21cm observations will enable imaging of reionization on very large scales. These images will contain more astrophysical and cosmological information than the power spectrum, and hence providing an alternative way to constrain the contribution of different reionizing sources populations to cosmic reionization. Using Convolutional Neural Networks, we present a simple network architecture that is sufficient to discriminate between Galaxy-dominated versus AGN-dominated models, even in the presence of simulated noise from different experiments such as the HERA and SKA.
\end{abstract}

\firstsection 
\section{Introduction}

The Epoch of Reionization (EoR) marks a period in the early universe, during which the birth of first luminous cosmic objects gradually reionized the neutral hydrogen in the Intergalactic medium (IGM). Studying this epoch will reveal a wealth of astrophysical and cosmological information concerning the nature of these first objects, and provide crucial input for theories of galaxy formation and evolution. Many ongoing and upcoming radio interferometer experiments such as the Low Frequency Array (LOFAR), the Precision Array for Probing the Epoch of Reionization (PAPER), the Hydrogen Epoch of Reionization Array (HERA), the Murchison Widefield Array (MWA), the Giant Metrewave Radio Telescope (GMRT), and the Square Kilometer Array (SKA), are promising to detect reionization in the near future through its 21cm fluctuations on large cosmological scales ($\gtrsim$ 500 Mpc). These experiments will provide a large amount of 21cm maps that encode these information. It is important to develop efficient statistical tools to best extract such information from upcoming 21cm survey data. 

The nature of sources driving cosmic reionization remains most uncertain. It has long been believed that star-forming galaxies have provided the full ionising photon budget required to complete reionization (e.g. \cite[Shapiro \& Giroux 1987; Hopkins et al. 2007]{shgi87,hop07}). However, several recent observational developments have triggered a debate about the ability of the Active Galactic Nuclei (AGN) to reionize the universe. These developments include \cite{gia15} high-redshift AGN observations, the flat slope of ionising emissivity measurements by \cite{bec13}, the early and extended Helium reionization by \cite{wor16}, and the large scale opacity fluctuations in the Ly$\alpha$ forest measured by \cite{bec15}.

In \cite{hassan18}, we have estimated the AGN contribution to reionization using a semi-numerical model, to which we add a plausible AGN contribution drawn from the \cite{gia15} observed luminosity function at z=5.75, and evolved to higher redshifts following a fixed Quasar Halo Occupancy Distribution. We have concluded that AGN-only models cannot simultaneously match current reionization constraints, namely the \cite{planck16} optical depth, ionising emissivity measurement by \cite{bec13}, and \cite{fan06} neutral fraction constraints by end of reionization. This indicates that AGN are highly unlikely to drive cosmic reionization, even if more faint AGN exist than previously thought. However, a model of 50\% contribution of AGN and galaxies barely matches these current constraints. This shows the need for additional set of observations which might likely provide more stringent constraints on AGN contribution to reionization. For this reason, we have performed 21cm forecasting for future observations by LOFAR, HERA and SKA and found that the 21cm power spectrum could potentially discriminate between these two models (\cite[Hassan et al. 2018]{hassan18}). 

Although the 21cm power spectrum is powerful in quantifying the large and small scale ionized bubbles clustering, the actual 21cm maps will contain more information to constrain the contribution of different reionization scenarios more efficiently. We next would like to assess the viability of using the 21cm maps to discriminate between AGN-only versus Galaxy-only models.

\section{Simulations}

\begin{figure}[b]
\begin{center}
\includegraphics[scale=0.36]{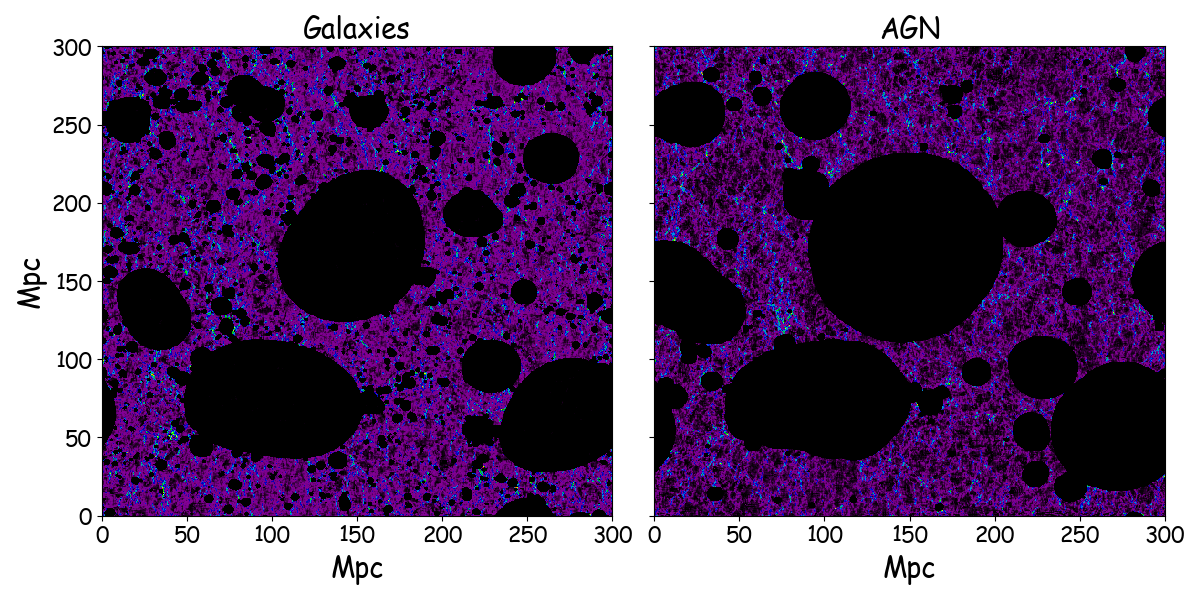} 
 \caption{The 21cm  brightness temperature at z=8 when reionization is half way through for Galaxies-only versus AGN-only models. Galaxies produce more small ionised bubbles structures while AGN produce larger and more spherical ionised bubbles due to the strong AGN clustering and high ionising emissivity ouput as implied by \cite{gia15} observations.}
   \label{fig1}
\end{center}
\end{figure}

We use an improved version of our semi-numerical code {\sc SimFast21} (\cite[Santos et al. 2010]{san10}) that has been recently developed in \cite{hassan17}. {\sc SimFast21} simulation begins by generating the dark matter density field using a Monte-Carlo Gaussian approach. The density field is then dynamically evolved into the non-linear regime via the Zel'dovich approximation. The dark matter halos are generated using the well known excursion set formalism (ESF, \cite[Press \& Schechter 1974]{prsc74}, \cite[Bond et al. 1991]{bond99}). In this improved model, the ionised regions are identified using a similar form of the ESF that is based on comparing the time-integrated ionisation rate with that of the recombination rate and the local neutral hydrogen density within each spherical volume specified by the ESF. Ionising photons from galaxies are modelled using a derived parametrization taken from a high-resolution radiative transfer simulation (\cite[Finlator et al. 2015]{fin15}) and a larger volume hydrodynamic simulation (\cite[Dav\'e et al. 2013]{dav13}), in order to account for the non-linear dependence on halo mass and redshift (see~\cite[Hassan et al. 2016]{hassan16} for more details on this derived parametrization and its effect on the 21cm signal). Ionising photons from AGN are computed by extrapolating the strong correlation between black hole mass and halos circular velocity from the local Universe observations (\cite[Ferrarese 2002]{fer02} and \cite[Tremaine et al. 2002]{tre02}). The AGN number density, and the corresponding duty cycle, are accounted for by adopting a fixed Quasar Halo Occupancy Distribution based on \cite{gia15} luminosity function. 
\begin{figure}[b]
\begin{center}
\includegraphics[scale=0.47]{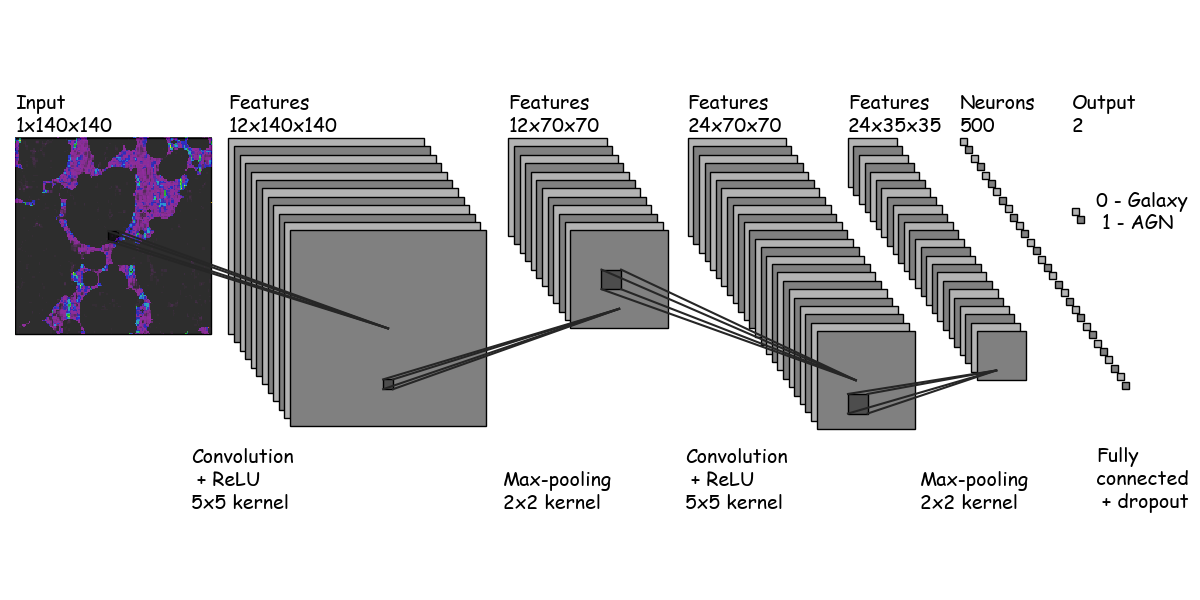} 
 \caption{Simple Convolutional Neural Network architecture used in the reionization models classifier. Each 21cm map is processed by 2 convolutional layers followed by pooling layers to eventually output the predicted model class (0 - Galaxy, 1 - AGN).}
   \label{fig2}
\end{center}
\end{figure}

\section{Reionization Models Classifier}
To classify between AGN-dominated and Galaxy-dominated models, we start by preparing our training and testing samples. The training set includes $\sim$ 10$^{3}$ 21cm images of 140$\times$140 pixels for each model from a simulation box of 75 Mpc. These 21cm images are taken out of different realizations by varying these models' free parameters, namely the photon escape fraction, ionising emissivity amplitude, mass power dependence. We account for the density field evolution by including images from several redshifts in the range z = 10 - 7 and a neutral fraction range of $x_{\rm HI}$ = 0.95 - 0.05. Figure~\ref{fig1} shows examples of the 21cm maps from our two models at fixed redshift (z = 8) and neutral fraction ($x_{\rm HI}$ = 0.5). As seen in this Figure, Galaxies models produce more small scale ionised bubbles while AGN-only models produce larger ionised bubbles. We next employ the LOFAR, HERA, and SKA final proposed configurations to add realistic random noise to the image space using {\sc 21cmSense} (\cite[Pober et al. 2014]{pob14}); a package for determining experiments' sensitivities to 21cm power spectrum. We use 90\% of those images to train our classifier and 10\% for validation purpose. 
\begin{figure}[b]
\begin{center}
\includegraphics[scale=0.37]{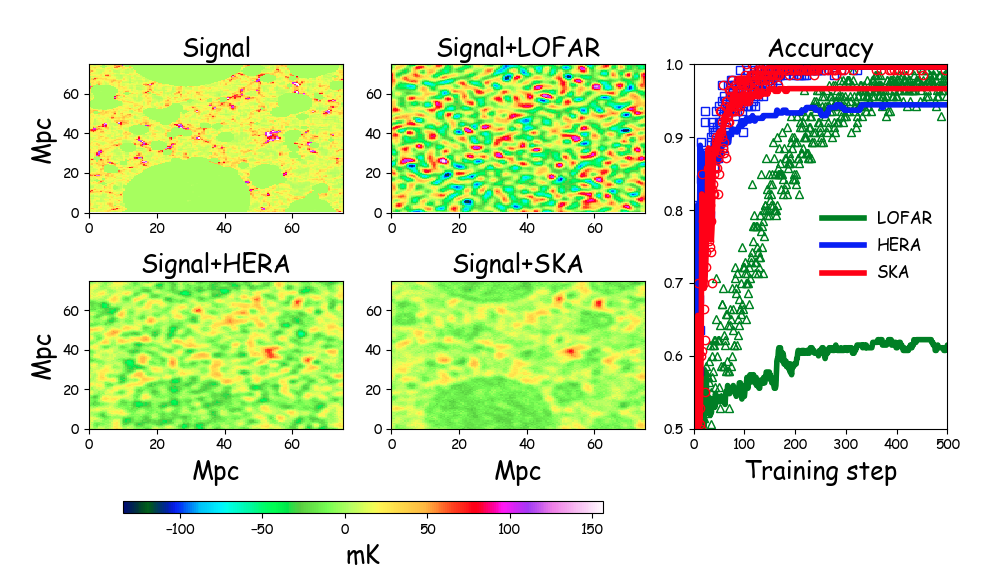} 
 \caption{Reionization Models Classifier using 21cm maps Deep Learning. Left: Realization of the 21cm signal from our AGN-dominated model with a simulated noise as expected from LOFAR, HERA and SKA. The training set contains $\sim$ 10$^{3}$ different realizations taken from our Galaxy-dominated and AGN-dominated models across several redshifts with different neutral fractions. Right: Classifier accuracy as a function of the training step for different experiments. Open symbols: green triangles (LOFAR), blue squares (HERA), and red circles (SKA) represent the accuracy of our training sample, while solid lines show the testing accuracy. The developed classifier is able to correctly recognize > 90\% of the training and testing samples from the  first 100 training steps. Such classifier can be used to constrain the contribution of different source populations during cosmic reionization from the future observations.}
   \label{fig3}
\end{center}
\end{figure}
We employ Convolutional Neural Networks (CNNs) in order to take to advantages of the 2-dimensional (2D) features (large versus small scale bubbles) encoded in the 21cm maps. Our CNN architecture is as follows: each 21cm map is first processed by a convolutional layer with 5$\times$5 filter of neurons batches, that generates 12 21cm features out of the input map. We then apply an 2$\times$2 max pooling layer to reduce their size from 140$\times$140 to 70$\times$70 images. This output is then processed by a similar convolutional layer to generate 24 more features, followed by a second pooling layer to reduce their size to half in pixels. We then process these 24 features of 35$\times$35 pixels into a fully connected layer to produce 500 different features. We apply the rectified linear unit (ReLU) activation function on all layers. To prevent overfitting and reduce the CNN complexity, we apply dropout on the fully connected layer to keep only 75\% of the neurons. Finally, these features are processed into an output layer with two neurons to obtain our predictions: 0 - Galaxy, 1 - AGN, as illustrated in Figure~\ref{fig2}. Figure~\ref{fig3} shows one realization of the 21cm maps from our AGN-only model, with a simulated noise added from different experiments. We find that the noise from LOFAR dominates the 21cm image, due to the low Signal-to-Noise ratio (SNR) and the few number of antennae. Unlike the case with LOFAR, the signal features are clearly seen in Figure~\ref{fig3} in the presence of noise from SKA and HERA, due to the high SNR and the large number of stations. In the same Figure, we display the classifier accuracy as a function of the training step. Open symbols: green triangles (LOFAR), blue squares (HERA), and red circles (SKA) represent the accuracy of our training sample, while solid lines show the testing accuracy. The developed classifier is able to correctly recognize > 90\% of the training and testing samples from the first 100 training steps. It is evident that this simple classifier is able to discriminate between our models in terms of their 21cm simulated maps, even in the presence of a realistic simulated noise as expected from future HERA and SKA observations.
\section{Summary}
We have presented a simple Convolutional Neural Network that is sufficient to discriminate between Galaxy-only and AGN-only models based on their 21cm maps as will be seen from HERA and SKA observations. This confirms the capability of using the actual 21cm maps as an alternative tool, besides the power spectrum, in order to efficiently handle the large amount of data expected from future 21cm observations. Our developed classifier can be used to constrain the contribution of different source populations during cosmic reionization from the future 21cm observations. 

It is worthwhile to mention that our training and testing samples are obtained by varying the astrophysical parameters, but assuming a fixed realization of the underlying density field. With these ideal samples, it is expected to obtain high accuracy with such a simple network. Realistically, our samples should include different density field realizations by changing the initial seed fluctuations. This step is currently under progress. Our future analysis will involve training the classifier on samples whose AGN-only and Galaxy-only models produce similar 21cm power spectra, and hence assessing the capability of using 21cm maps in extreme cases when the power spectrum is kept fixed. Using the final trained architecture, we will be able to identify the features (e.g. large versus small bubbles) by which the classification is determined. We finally plan to introduce an unrealistic third model, in which the correlation between the density and ionization fields is broken. This third model will test the classifier ability to discriminate between a mixture of realistic and unrealistic reionization models. All these steps are required to add more complexity and generalize our analysis in order to provide a more generic robust classifier for the upcoming 21cm surveys.

\end{document}